\documentclass[prl,reprint,longbibliography]{revtex4-1}

\usepackage{graphicx}
\usepackage{amssymb}

\newcommand{\kms}{\ensuremath{\mathrm{km}\,\mathrm{s}^{-1}}}

\newcommand{\LCDM}{$\Lambda$CDM}
\newcommand{\Lnoob}{NoCDM}
\newcommand{\LyA}{\ensuremath{\mathrm{Ly}\alpha}}
\newcommand{\Trad}{\ensuremath{T_{\gamma}}}
\newcommand{\Tgas}{\ensuremath{T_{k}}}
\newcommand{\Tspin}{\ensuremath{T_{S}}}
\newcommand{\Tb}{\ensuremath{T_{21}}}

\begin{document}

\title{Predictions for the sky-averaged depth of the 21cm absorption signal at high redshift
in cosmologies with and without non-baryonic cold dark matter}

\author{Stacy S. McGaugh} 
\affiliation{Department of Astronomy, Case Western Reserve University, 10900 Euclid Avenue, Cleveland, OH 44106, USA}

\date{\today}

\begin{abstract}
We consider the 21cm absorption signal expected at high redshift in cosmologies with and without non-baryonic cold dark matter. 
The expansion of the early universe decelerates strongly with dark matter, but approximately coasts without it. 
This results in a different path length across the epochs when absorption is expected, with the consequence that
the absorption is predicted to be a factor of $\sim 2$ greater without dark matter than with it.
Observation of such a signal would motivate consideration of extended theories of gravity in lieu of dark matter.
\end{abstract}


\maketitle

\section{Introduction}

The nature of the missing mass remains one of the great unsolved problems in physics.
The existence of non-baryonic cold dark matter (CDM) is an apparent requirement of modern cosmology for a variety
of reasons \cite{Peebles}, perhaps most notably \cite{WMAP3,Planck15} the cosmic microwave background (CMB).
This cosmic dark matter is widely assumed to be a weakly interacting massive particle (WIMP),
yet decades of direct detection experiments have so far yielded only null results \cite{Akerib,PandaX,Xenon1T}. 
These non-detections have repeatedly excluded regions of parameter space where positive detections had been expected \cite{Trotta}.
More generally, there is no positive experimental evidence for supersymmetry, a necessary prerequisite for the hypothesized WIMPs.
Meanwhile, some astronomical data provide reason to doubt the existence of CDM outright \cite{sandersNOCDM,Kroupa2012,MdB98a}.

This situation motivates consideration of alternatives to WIMP dark matter.
Here we consider the possibility that the effects we have been interpreting as dark matter might in fact point to a need for an extended theory of gravity.
Such a possibility is observationally well motivated \cite{SMmond,FM12}, with powerful arguments that can be made for and against both 
the dark matter and extended gravity interpretations \cite{CJP}. 

The modified Newtonian dynamics
(MOND) hypothesized by Milgrom \cite{MONDorig} has had many a priori predictions \cite{milgrom83} come 
true \cite{BBS,MdB98b,PRL11,MM13a,MM13b,RAR,LMSP,LLMS18}. This should not happen if dark matter is the correct interpretation
of the observed discrepancies. However, while MOND has the interesting property that 
dynamics \footnote{Though it has become conventional to discuss modifications of gravity, MOND might also be interpreted as a modification of inertia.} 
become scale invariant \cite{scaleinvar}, 
attempts to incorporate it into a generally covariant framework have been frustrating. The most prominent example of such a theory, TeVeS \cite{TeVeS},
fails to fit the CMB \cite{TVSforcing}, grow structure \cite{Dod2011}, or be consistent with observations of gravitational waves \cite{SandersGW}.

The failure of the specific theory TeVeS does not falsify the more general hypothesis of scale invariant 
dynamics \footnote{Dynamics in the deep MOND regime are scale invariant under transformations $(t, \mathbf{r}) \rightarrow (\lambda t, \lambda \mathbf{r})$}, 
but we are left without a clear theory for cosmology in this context. 
Fortunately, the physics of the 21cm absorption is straightforward, depending only on atomic physics and the fact that the universe is expanding.
This provides the opportunity to outline some very general expectations. Generically, we expect a universe devoid of non-baryonic dark matter (\Lnoob)
to be low density, and thus experience less deceleration at early times \cite{Felten} than the 
conventional \LCDM\ \footnote{One might interpret the need to invoke both dark matter and dark energy as a failing of current theory:
these are auxiliary hypotheses invoked to save the phenomena of a Friedmann-Robertson-Walker cosmology.}  
universe. As a consequence, there is a greater path length to the surface of last scattering that leads to a stronger absorption signal.

\section{21cm Absorption in the Early Universe}

The early ($10 \lesssim z \lesssim 1000$) universe contains an enormous amount of information \cite{LZ2004}, but remains largely unexplored. 
These epochs follow recombination but precede the bulk of star formation, so the baryonic content is expected to be a largely neutral gas composed of 
hydrogen and helium in their primordial 3:1 mass ratio \cite{BBNorig,BBN}. During this time \footnote{The age of the universe 
ranges from a few $10^5$ yr at recombination to 20 Myr at $z=100$ (the dark ages) to 180 Myr at $z=20$ (cosmic dawn) in \LCDM.}, 
we expect an absorption signal from the 21cm spin-flip transition
of neutral hydrogen \cite{PL12} seen against the backdrop provided by the CMB from $z = 1090$ \cite{Planck15}.

The relevant atomic physics is straightforward yet remarkably rich. 
We are concerned with three distinct temperatures: that of the radiation field \Trad,
the kinetic temperature of the gas \Tgas, and the spin temperature of the 21cm line \Tspin\ that specifies the occupation of the hyperfine levels \cite{PL12}.
The expansion of the universe leads to a divergence between the kinetic temperature of the gas, which varies approximately as $(1+z)^2$, 
and that of the radiation field, which varies as $(1+z)$. 
The spin temperature is bounded between the two, $\Tgas \le \Tspin \le \Trad$,
and specified by \cite{PL12}
\begin{equation}
\Tspin^{-1} = \frac{\Trad^{-1} + x_i \Tgas^{-1}}{1+x_i}.
\label{eq:spinT}
\end{equation}
Here $x_i$ generically represents any physical effect that couples the spin temperature to the gas.
We expect absorption whenever $\Tspin < \Trad$.

There are two distinct physical processes at work at different points in the early universe.
Consequently, we expect to see 21cm absorption against the microwave background at two distinct epochs, 
`cosmic dawn' ($z \approx 20$) and the `dark ages' ($z \approx 100$). 
During the dark ages, the dominant coupling $x_i$ is collisional.
At cosmic dawn, the dominant mechanism is the Wouthuysen-Field effect \cite{Wout,Field58} stemming from the resonant scattering 
of \LyA\ photons produced by the first stars.

To compute the spin temperature during the dark ages, we utilize existing fitting functions for $x_i$: see equation 10 of \cite{PL12} and the subsequent text. 
As the universe expands, the spin temperature begins to diverge from the radiation temperature after decoupling, tending towards the cooler gas
temperature. This leads to the expectation of an absorption signal that is maximized around $z \approx 100$, after which the spin temperature
reverts to the radiation temperature as collisions become rare in the expanding universe. 
This is straightforward atomic physics driven by H-H and e-H scattering, so provides an especially clean test.

As collisions decline, the Wouthuysen-Field effect begins to dominate the coupling.
This is driven by the appearance of the first stars that produce \LyA\ photons.
Resonant coupling with these \LyA\ photons plays the same role as collisions during the dark ages.
The spin temperature again diverges from the radiation temperature, approaches the gas temperature, then rebounds,
giving the expectation of another redshift dependent absorption trough \cite{PL12}. 

Proper estimation of the Wouthuysen-Field effect depends on the spectrum of the first stars, and thus
requires the construction of a model for these first sources of radiation \cite{ZFH,FP06,PL12}.
Such models are, of necessity, rather uncertain, though we note that in most plausible models, the coupling is efficient
thanks to the high optical depth of \LyA\ photons so that the spin temperature very nearly reaches
the gas temperature \cite{CFBL}. The maximum possible absorption signal occurs in the limit $\Tspin \rightarrow \Tgas$, so rather than
build a specific model, we take the general case that gives the maximum possible absorption signal.
This follows simply by setting $\Tspin = \Tgas$ at the redshifts of interest ($16 \le z \le 19$). 
The actual signal might be a bit less, but it certainly should not be more in a given cosmology.

Quantitatively, the amplitude \cite{ZFH,2018RNAAS} of the 21cm brightness temperature \Tb\ as a function of redshift $z$ is given by
\begin{equation}
\Tb(z) = T_0\, \frac{\mathrm{x}_{\mathrm{HI}}}{\mathfrak{h}_z} \left[ (1+z) f_b \left( \frac{\omega_b}{0.02} \right) \right]^{1/2} \left(1- \frac{T_{\gamma}}{T_S} \right).
\label{eq:T21}
\end{equation}
Here $T_0 = 20\; \mathrm{mK}$, $\mathrm{x}_{\mathrm{HI}}$ is the neutral hydrogen fraction (very nearly unity over the relevant range of redshifts), 
$\mathfrak{h}_z$ is a factor for the cosmology-specific expansion history defined below, 
$f_b = \Omega_b/(\Omega_b+\Omega_{\mathrm{CDM}})$ is the cosmic baryon fraction, 
and $\omega_b = \Omega_b h^2$ \footnote{$h = H_0/(100\;\kms\,\mathrm{Mpc}^{-1}$).}
is the baryon density.
Three basic elements impact the optical depth of the 21cm line seen in redshifted absorption towards the CMB:
the temperature evolution of the universe, the density of hydrogen atoms, and the path length along which the absorption occurs.

The evolution of the radiation temperature is simply $\Trad = (2.725\;\mathrm{K}) (1+z)$. The evolution of the gas temperature requires numerical
solution of the Saha equation in an expanding universe, which we accomplish with the industry-standard 
code RECFAST \cite{RECFAST1,RECFAST2}, including all updates to the atomic physics \cite{RECFAST3,RECFAST4}.
RECFAST also provides the neutral fraction of hydrogen; the cosmologies considered here have a very similar temperature history 
and $\mathrm{x}_{\mathrm{HI}}(z)$.

\section{Cosmological Models}

To illustrate the expected difference between a universe with and without non-baryonic dark matter, we compute the temperature evolution and
corresponding 21cm absorption signal for two cosmologies: \LCDM\ and \Lnoob\ ($\Omega_{\mathrm{CDM}} = 0$). 
We adopt for \LCDM\ parameters from Planck \cite{Planck15} for input to RECFAST as noted in Table \ref{tab:T21}.
For eq.\ \ref{eq:T21}, the relevant quantities are $\omega_b = 0.022$ and $f_b = 0.156$. 
These parameters are now so tightly constrained that the absorption signal from the dark ages must follow with little uncertainty, and that
at cosmic dawn is strongly bounded.

For specificity, we adopt the \Lnoob\ model from the last row of Table 3 in \cite{M2004CMB}.
However, most of the parameters explored in detail there are irrelevant here.
We have calculated variations on the \Lnoob\ model; the results presented here hold generically provided 
that $\omega_b \approx 0.02$ and $\Omega_{\mathrm{CDM}} = 0$, irrespective of other cosmological 
parameters \footnote{The specific value of $\Omega_{\Lambda}$ and many other details are not very important for
this problem at high redshift. It is worth noting that a coasting universe, while consistent with a good deal of data,
cannot explain the location of the first acoustic peak in the CMB. This may be a clue that the usual definition of the angular diameter 
distance may differ in the underlying theory: the geometry must be near flat in the Robertson-Walker sense while the expansion
might resemble the $\Omega \approx 0$ case.}
 
The great success of \Lnoob\ is that it correctly predicted \cite{M1999CMB} the amplitude of the 
second peak \citep{M2000} in the acoustic power spectrum of the CMB, and was the only model to do so a priori.
The chief failing of \Lnoob\ is that it underpredicts the amplitude of the third and subsequent 
peaks \footnote{This aspect of the CMB data is widely considered to falsify MOND and all extended theories
of gravity. This is an over-interpretation, as only the ansatz on which \Lnoob\ is based is falsified, as it must be at some point. 
All we really know so far is what does not work: General Relativity plus known particles. 
The need for new particles beyond those of the Standard Model follows absolutely only after we assume General Relativity.}. 
Despite preceding the Planck satellite by a decade, the specific model of \cite{M2004CMB} utilized here
provides an excellent fit \cite{CJP} to the Planck data with $\ell < 600$ with zero adjustment.
However, it provides no fit nor explanation for the data at $\ell > 600$, and we are left to imagine that some feature of a more general
theory might account for this aspect of the data. 
The forcing term we currently attribute to CDM might instead be caused, e.g., by
a scalar field \cite{TVSforcing}, but I am not aware of an extended theory of gravity that provides a detailed fit to the acoustic
power spectrum in the same sense that \LCDM\ does.
 
For our purposes here, \Lnoob\ is merely a proxy for the unknown unknowns.
The important difference from \LCDM\ is the baryon fraction, $f_b = 1$. We adopt $\omega_b = 0.022$ to be identical with \LCDM\ so that
the baryon fraction is the main difference in eq.\ \ref{eq:T21}. Notably, to maintain $\Omega_b = 0.039$ as used in \cite{M2004CMB} implies
$h = 0.75$: the Hubble constant is thus not in tension with local measurements \cite{CosmicFlows3,RiessH0}.

\begin{table}
\caption{\label{tab:T21} Predicted Spin Temperatures and 21cm Absorption}
\begin{ruledtabular}
\begin{tabular}{|r|c|c|c|c|r|r|}
 &  &  & \LCDM$^{\mathrm{a}}$ & \Lnoob$^{\mathrm{b}}$ & \LCDM\ & \Lnoob\ \\
\hline
$z$ & $\nu$ (MHz) & \Trad\ (K) & \multicolumn{2}{c|}{\Tspin\ (K)} & \multicolumn{2}{c|}{\Tb\ (mK)} \\
\hline
\multicolumn{7}{|l|}{Cosmic Dawn} \\
\hline
16 & 83 & 46.3 & 6.16 & 6.56 & $-226$ & $-499$ \\ 
17 &Ê79 & 49.1 & 6.90 & 7.34 & $-218$ & $-482$ \\
18 & 75 & 51.8 & 7.67 & 8.15 & $-211$ & $-467$ \\
19 & 71 & 54.5 & 8.48 & 9.02 & $-204$ & $-452$ \\
\hline
\multicolumn{7}{|l|}{Dark Ages} \\
\hline
50 & 28 & 139 & 119 & 119 & $-10$ &  $-22$ \\
100Ê& 14 & 275 & 197 & 203 & $-33$ & $-68$ \\
200 & ~7 & 548 & 471 & 478 & $-19$ & $-37$ \\
\end{tabular}
\end{ruledtabular}
\footnotetext{\LCDM: $\Omega_b = 0.0488$, $\Omega_{\mathrm{CDM}} = 0.2633$, $\Omega_{\Lambda} = 0.6879$, $h = 0.675$.}
\footnotetext{\Lnoob: $\Omega_b = 0.039$, $\Omega_{\mathrm{CDM}} = 0$, $\Omega_{\Lambda} =  0.91$, $h = 0.75$ (RECFAST inputs).}
\end{table}

The way eq.\ \ref{eq:T21} is written subsumes the expansion history of the universe.
This is critical, as the path length for absorption depends on the variation of the Hubble parameter, $H(z)$. 
Eq. \ref{eq:T21} has been derived \cite{ZFH} using the approximation 
\begin{equation}
\tilde H(z) = H_0 \Omega_m^{1/2} (1+z)^{3/2}.
\label{eqn:approxHz}
\end{equation}
This suffices for \LCDM\ but is inadequate for \Lnoob, so we compute for both the full expression
\begin{equation}
H^2(z) = H_0^2 [\Omega_{\Lambda}+\Omega_m(1+z)^3+\Omega_r(1+z)^4-\Omega_k(1+z)^2]
\label{eqn:Hz}
\end{equation}
where $\Omega_r$ is the radiation density, including neutrinos. This is not entirely negligible in \Lnoob\ at the higher redshifts of interest here.
Together, equations\ \ref{eqn:approxHz} and \ref{eqn:Hz} provide the correction factor $\mathfrak{h}_z = H(z)/\tilde H(z)$. 
It is at most a few percent for \LCDM. For \Lnoob, it is 5\% at
cosmic dawn, and $\sim 10\%$ during the dark ages ($z \approx 100$). While the specific \Lnoob\ model we adopt provides the 
run of $\mathfrak{h}_z$ necessary for the calculation here, more generally the absorption signals provide an empirical constraint 
on the expansion history $H(z)$ \cite{MM2015,WMW2017,Maeder2017} in a huge but otherwise inaccessible volume of the universe.
A significant deviation from the predictions of both \LCDM\ and our simple \Lnoob\ model would be a strong hint that the expansion
history of the universe is incompatible with standard cosmology.

\begin{figure*}
\includegraphics[width=6.0in]{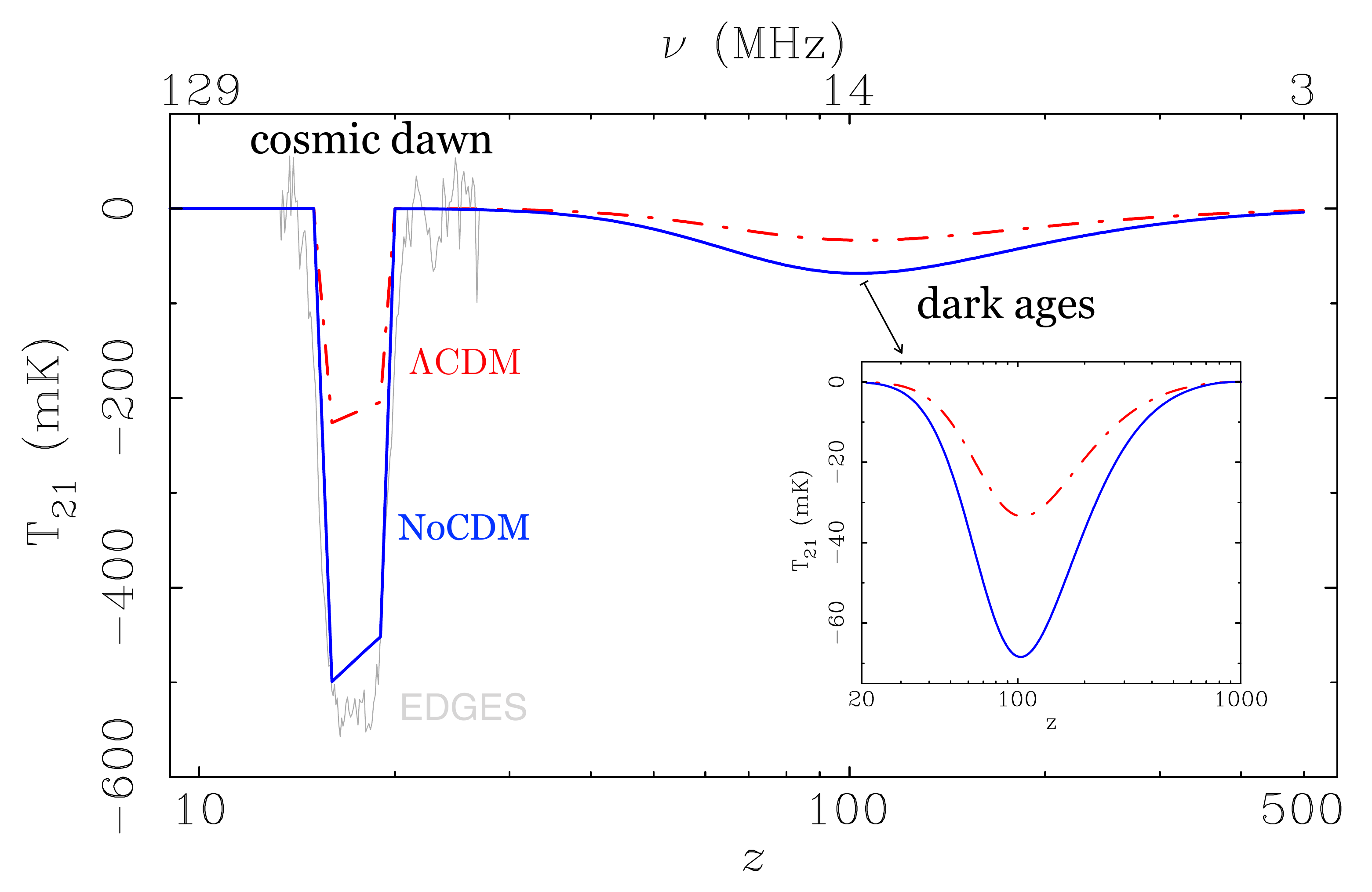}
\caption{The sky-averaged 21cm absorption as a function of redshift, with the corresponding frequency noted at top. 
The prediction for \LCDM\ with Planck parameters \cite{Planck15} is shown as the broken red line; 
a model (\Lnoob\ \cite{M2004CMB}) without cold dark matter is shown as the solid blue line. 
The inset magnifies the absorption during the dark ages, which is expected to be 
weaker than that at cosmic dawn. The dark ages provide a cleaner test, as the signal at cosmic dawn depends on the spectrum produced by the first stars.
We ignore these astrophysical details here to show the maximum possible signal by simply setting $T_s = T_g$ for $16 \le z \le 19$. This redshift range is
motivated by the detection reported by EGDES \cite{EDGES}, shown in gray.
\label{fig:T21}}
\end{figure*}

\section{Results}

The main difference in eq.\ \ref{eq:T21} between the cosmologies considered here is the baryon fraction: $f_b = 0.156$ in \LCDM\ and $f_b = 1$ in \Lnoob.
This leads to a difference in
absorption signal, as illustrated in Table \ref{tab:T21} and Fig.\ \ref{fig:T21}. One expects roughly twice as much absorption in \Lnoob\ as in \LCDM.

Physically, the reason for the greater optical depth in \Lnoob\ is a greater path length to the surface of last scattering, not an increase in the density of
hydrogen. The key factor encoded in $f_b$ in eq.\ \ref{eq:T21} is the difference in mass density $\Omega_m$ in eq.\ \ref{eqn:approxHz}. A universe with
a large mass density like \LCDM\ decelerates strongly at the high redshifts considered here before accelerating at late times. A low density universe
devoid of non-baryonic dark matter is close to the coasting limit. This leads to a longer path length and correspondingly greater optical depth.

We must at this juncture bear in mind that \Lnoob\ is merely a proxy for some unknown, underlying theory. This could be quite different, though hopefully
distinct from \LCDM.  However, the near-coasting limit seems appropriate for a very simple physical reason. Without dark matter, there  
isn't much mass for gravity to operate on, even if the force law is modified. Consequently, the near-coasting approximation should 
generically be a good one, as it is in the specific case of MOND \cite{Felten,Sanders98}.

The expected absorption at cosmic dawn has recently been detected by EDGES \cite{EDGES}. This is shown in Fig.\ \ref{fig:T21} along
with the model predictions. The depth of the observed absorption is consistent with \Lnoob, within the formal uncertainties. The possible
absorption ranges over $451 < |\Tb| < 499$ mK compared to the observed $500^{+500}_{-200}$ mK at 99\% c.l. \cite{EDGES}. 
The predicted range is for the maximum possible absorption, which we expect to be very nearly realized in practice \cite{CFBL}.

In contrast, \LCDM\ is not consistent with EDGES \cite{EDGES,DMsanetalk,lightDMcrazytalk}. 
For Planck parameters, the maximum possible absorption is 226 mK at $z=16$ (Table \ref{tab:T21}). 
This is significantly less than the 99\% confidence limit of $|\Tb| > 300$ mK \cite{EDGES}. 

Preserving \LCDM\ requires some form of special pleading. Examination of eq.\ \ref{eq:T21} shows that a signal of arbitrary strength may be
obtained by adjusting the ratio $\Trad/\Tspin$. These are set by cosmology and atomic physics, so are not easily altered.
Nevertheless, the necessary effect might be obtained by increasing \Trad\ with some hypothetical radio sources at
very high redshift \cite{RadioBGDT,RadioBG21}, or by decreasing \Tspin\ with some non-standard physics \cite{Barkingmad,FBC18,BHKM18,HB18}. 
These are unnatural auxiliary hypotheses of the sort that can always be invoked to save the phenomena. 
In contrast, the observed signal is entirely natural in a universe devoid of CDM \cite{M1999fb,2018RNAAS}.

The EDGES \cite{EDGES} result is the first detection of the expected absorption, and may be subject to a variety of systematic errors.
While we have no reason to doubt its veracity, it is a challenging observation that requires subtraction of strong foreground signals.
Given the importance of the result, we hope to see it independently checked by other experiments. 

So far, we have considered only the magnitude of the absorption. There is further information in its timing and shape.
Given the tentative nature of the detection, we caution against over-interpreting these features. For completeness, we note
that the early occurrence of cosmic dawn
is natural in MOND \cite{Sanders98}, though it may also be obtained in \LCDM\ \cite{KVZD18} depending on the nature of the first sources 
(see \cite{M2004CMB} and references therein). The shape of the signal is entirely unexpected, being much sharper than anticipated \cite{MF18}.
The simple ``on-off" model adopted here provides a surprisingly good match to the observed redshift dependence without attempting
to do so. This might be suggestive of the sudden onset of structure formation anticipated in MOND \cite{Sanders98}.

\section{Further Predictions}

\subsection{Dark Ages}

The EDGES \cite{EDGES} detection at cosmic dawn opens a new window on the early universe.
At still higher redshift, the dark ages remain to be probed. The prediction of \Lnoob\ is that the amplitude of absorption will again be about twice
that expected in \LCDM\ with Planck \cite{Planck15} parameters (Table \ref{tab:T21}). This is in principle a clean test, because simple
atomic scattering dominates the coupling of \Tspin\ to \Tgas. Both are readily calculable, and depend only on atomic physics and the fact that
the universe is expanding. 

In order to observe the signal at $z \approx 100$ requires very low frequency data, below the Earth's ionospheric cut-off ($\nu < 30$ MHz). 
Such data would be best obtained from the far side of the moon, using the moon to shade the instrument from terrestrial radio interference. 
These results add motivation for such an experiment \cite{Farside13}. Detecting the 21cm signal from the dark ages is already compelling motivation,
but it additionally enables a fundamental test of these differing cosmological hypotheses.

\subsection{Structure Formation}

The discussion above is as generic as possible, seeking only a test of whether the universe contains CDM
(see also \cite{Kroupa2012,Kroupa2014}). 
The approach here suffices to consider the sky-averaged 21cm absorption intensity. 
However, we also anticipate fluctuations in this intensity along different lines of sight \cite{PL12}. 
I discuss here some basic expectations for the power spectrum of these fluctuations.

Following the MOND cosmogony outlined by Sanders \cite{Sanders98}, no growth can occur prior to 
decoupling from the radiation field at $z \sim 200$. Once the radiation loses its grip, density fluctuations find themselves deep in the low acceleration regime. 
Consequently, the MOND effect is large, and they behave as if there were lots of dark matter: structure formation proceeds 
rapidly \cite{Sanders01,SK01,Nusser02a,Llinares08}. The Sanders cosmogony anticipates large
galaxies forming at $z \approx 10$, clusters of galaxies at $z \approx 3$, and the emergence of the cosmic web at $z \sim 5$. In contrast,
contemporaneous \LCDM\ models predicted very little structure at these times, with only 1\% of stars forming at $z > 5$ \cite{Baugh98}.
The early onset of star formation \cite{Hashimoto2018}, the `impossibly early galaxy problem' \cite{impossiblyearly}, the presence of clusters of galaxies 
in the early universe \cite{CCPCII,CPCCSpitzer,z43cluster,z57cluster}, 
and the strong clustering of quasars at high redshift \cite{Hizquasarclustering} were all anticipated 
long before their observation by the Sanders cosmogony \cite{Sanders98}.

MOND is non-linear, so quantitative prediction of the power spectrum \cite{Sanders01} is beyond the scope of this note.
Nonetheless, we can anticipate the amount of power relative to the linear growth of perturbations in \LCDM. 
Early on, $z \approx 200$, we expect there to be less power than in \LCDM\ due to Silk damping. 
By $z \approx 30$, this situation will have reversed, with structures
forming rapidly over a large range of mass scales, yielding more fluctuation than expected in \LCDM. 

The rapid development of structure at high redshift anticipated in the Sanders cosmogony \cite{Sanders98,SK01} leads to the expectation
that the formation of the first stars is more sudden and widespread than expected in \LCDM \cite{M2004CMB}. This leads to early reoinization:
the optical depth to the surface of last scattering is $\tau \approx 0.17$ in \cite{M2004CMB}. 
The sudden onset of cosmic dawn is qualitatively consistent with the shape
of the EDGES absorption trough (Fig.\ \ref{fig:T21}). 

Another expectation for the power spectrum at high redshift is the presence of pronounced baryonic features \cite{M1999fb,Dod2011}.
These do not necessarily persist to low redshift because mode mixing is likely in the non-linear structure formation driven by MOND. 
However, the power spectrum that first emerges after decoupling should preserve the imprint of strong baryonic oscillations \cite{M1999fb}.
This is best tested during the dark ages at the highest accessible redshift.

\subsection{Neutrino Mass}

The range of allowed neutrino mass in \LCDM\ is strictly limited by neutrino oscillations \cite{neutrinomassorder} and
the latest Planck results \cite{Planck2018} to fall in the narrow range $0.06 < \sum m_{\nu} < 0.12$ eV. 
Though not relevant to 21cm absorption, there are indications of a larger neutrino mass ($m_{\nu} > 0.5$ eV) 
in the models considered by \cite{M2004CMB}. This provides another test to distinguish between \LCDM\ and \Lnoob. 
An experimental determination that $\sum m_{\nu} > 0.12$ eV would falsify \LCDM.
The same limit does not apply in MOND, for which $m_{\nu} \sim 1$ eV
might help prevent it from overproducing structure \cite{Nusser02a} and explain the excess discrepancy in clusters of galaxies \cite{sandersclusters}.

\section{Conclusions}

A challenge for modern cosmological theory is to identify definitive predictions by which  \LCDM\ would be subject to falsification \cite{Kroupa2012,Kroupa2014}.
Previous attempts \cite{M1999CMB} with the CMB succeeded \cite{M2000,M2004CMB} before failing \cite{CJP} due to previously unacceptable degrees of 
freedom \footnote{It was long believed that BBN required $\omega_b = 0.0125$ until fitting CMB data required $\omega_b = 0.0224$.
No BBN measurements prior to the first relevant CMB measurements had suggested $\omega_b > 0.020$}. 
I worry that the current picture allows so much room for auxiliary hypotheses that there is no possibility 
of discerning that it is incorrect, should that happen to be the case.

Here I have highlighted the amplitude of the 21cm absorption as a possible test. Generically, I predict that a universe devoid of CDM will
exhibit about twice as much absorption as is possible in \LCDM. Taken at face value, the EDGES \cite{EDGES} detection 
is in serious conflict with \LCDM\ \cite{DMsanetalk} while corroborating the \Lnoob\ calculation made here. 
The various auxiliary hypotheses \cite{lightDMcrazytalk} that have been offered to reconcile the EDGES signal 
with \LCDM\ highlight my concern for its falsifiability.

\medskip
\begin{acknowledgements}
I am grateful to Avi Loeb and his collaborators for their lucid reviews of the relevant physics, and to the referees for their constructive comments. 
\end{acknowledgements}


%

\end{document}